# Dithymoquinone as a novel inhibitor for 3-carboxy-4-methyl-

# 5-propyl-2-furanpropanoic acid (CMPF) to prevent renal

# Failure

# (that we are currently working on)

Muniba Faiza[1], Tariq Abdullah[2], Prof. Yonghua Wang[1*]

[1]School of Food Science and Engineering, South China University of Technology, Guangzhou 510640, China; Email: muniba.faiza@gmail.com

[2]IQL Bioinformatics, Idea Quotient Labs, New Delhi, India; Email: tariq@ideaquotient.in

*Corresponding Author

Yonghua Wang, Ph.D.

**Vice dean in the College of Light Industry and Food Sciences**
School of Food Science and Engineering,
South China University of Technology,
Guangzhou, Guangdong Province, China 450001
Tel and Fax: 86-020-87113842
E-mail: yonghwang@scut.edu.cn

# ABSTRACT

3-carboxy-4-methyl-5-propyl-2-furanpropanoic acid (CMPF) is a major endogenous ligand found in the human serum albumin (HSA) of renal failure patients. It gets accumulated in the HSA and its concentration in sera of patients may reflect the chronicity of renal failure [1-4]. It is considered uremic toxin due to its damaging effect to the renal cells. The high concentrations of CMPF inhibit the binding of other ligands to HSA. Removal of CMPF is difficult through conventional hemodialysis due to its strong binding affinity. We hypothesized that the competitive inhibition may be helpful in removal of CMPF binding to HSA. A compound with higher HSA binding affinity than CMPF could be useful to prevent CMPF from binding so that CMPF could be excreted by the body through the urine. We studied an active compound dihydrothymoquinone/ dithymoquinone (DTQ) found in black cumin seed (*Nigella sativa*), which has higher binding affinity for HSA. Molecular docking simulations were performed to find the binding affinity of CMPF and DTQ with HSA. DTQ was found to have higher binding affinity possessing more interactions with the binding residues than the CMPF. We studied the binding pocket flexibility of CMPF and DTQ to analyze the binding abilities of both the compounds. We have also predicted the ADME properties for DTQ which shows higher lipophilicity, higher gastrointestinal (GI) absorption, and blood-brain barrier (BBB) permeability. We discovered that DTQ has potential to act as an inhibitor of CMPF and can be considered as a candidate for the formation of the therapeutic drug against CMPF.

## KEYWORDS

CMPF, renal failure, black cumin seed (*Nigella sativa*), thymoquinone, dithymoquinone, molecular docking, binding affinity.

# INTRODUCTION

CMPF has been linked to the acute renal failure [1-3,5]. CMPF is a major endogenous toxin retained in uremic serum and major drug-binding inhibitor [6]. The accumulation of CMPF is caused by binding of CMPF to HSA, which cannot be removed through conventional hemodialysis [7]. Currently, there is no known inhibitor for CMPF which can reduce its accumulation.

## Our Hypothesis

CMPF accumulation is due to its high binding affinity towards HSA among uremic toxins [8-10]. We hypothesized that another compound with a higher binding affinity towards HSA than CMPF should be able to bind with HSA. Thus, effectively blocking CMPF from binding. This could have beneficial effects on human health apart from the effects on kidneys. The binding of CMPF could be reduced leading to the decreased chances of acute renal failure. Assuming this, we hypothetically considered a compound which has been earlier proved to be beneficial for human health, such as cancer, tumor, and more [11-17]. This compound is dithymoquinone (DTQ), one of the active compounds found in black cumin seed (*Nigella sativa*), which is well known for its positive effects on human in various conditions such as tumor, cancer, clearance of hepato-renal toxicity, diabetes, [8-18].and found to possess relatively higher binding affinity than CMPFWe predicted the inhibitory effect of DTQ against CMPF using *in silico* docking performed by AutoDock tools [19]. In order to determine the capability of DTQ as the potential drug candidate against CMPF, we predicted its ADME properties using SwissADME [20]. The crucial role of CMPF in renal failure, binding with HSA, and DTQ is discussed in detail in the following sections.

## HSA- receptor for endogenous substances and drugs

HSA is an abundant plasma protein that belongs to a multigene family of proteins including α-fetoprotein, vitamin-D binding protein, and human group specific component [21]. HSA binds to a wide range of drugs and/ or endogenous ligands including bilirubin, thyroxine, non-esterified fatty acids, and hemin [21], all of them being lipophilic, acidic compounds, in multiple sites [22-27]. HSA protein consists of 585 amino acids with 17 disulphide bridges, one tryptophan residue (Trp 214), and a single free-thiol (Cys 34) [28]. According to Sudlow et al., (1975) [29], there are two binding sites in HSA: Site I and Site II. He & Carter (1992) [28] observed the binding pockets of HSA lying in two subdomains: IIA and IIIA, which were the same sites but they used different nomenclature to denote them. HSA is considered a silent receptor for drugs [30] because it serves as a repository for a wide variety of compounds as the negative charge on HSA facilitates the electrostatic binding of various ligands and acts as a depot and carrier for many drug compounds [31-34]. Due to the presence of limited number of high- affinity binding sites in HSA, detailed molecular information about these sites could be helpful in studying the effects of other drugs or ligands and the structural information of both low- and high- affinity binding sites is useful in designing new drugs with the aim either avoiding binding to HSA or to make use of its functions [35].

CMPF is a metabolite of F-acids [36-40] which was first detected in human urine [37] and then later in human blood [38]. CMPF is considered as one of the major drug- binding inhibitor [1,6,41-43]. The drug-binding effect observed *in vivo* that the CMPF levels rise in kidney patients are due to the specific steric blocking of drug site 1 by this compound [43,44]. Sakai et al., (1995) [44] studied the interactions of CMPF in comparison of other three uremic toxins: indoxyl sulphate, indole acetic acid, and hippuric acid using fluorescent probe displacement, ultrafiltration, and equilibrium dialysis. The results showed that CMPF had the highest binding affinity ($10^{-7}$ $M^{-1}$) among the three uremic toxins. As compared with the other uremic toxins, CMPF consists of least rate of elimination and the most potent inhibitor [10,44-47]. Several studies have demonstrated that uremic toxins are involved in the production of reactive oxygen species (ROS) [48-50]. In a study conducted by Miyamoto et al., (2012) [50], it was shown that CMPF induces the renal cell damage by enhancing the production of ROS in the presence of Angiotensin-II (A-II), which is an inducer of oxygen free radicals. Also, in the presence of iron, CMPF and A-II induce the Fenton's reaction leading to an increase in ROS production. The subsequent interaction of CMPF with dissolved oxygen leads to the overproduction of oxygen free radicals.

All the experiments mentioned above are sufficient enough to prove the damaging role of CMPF in renal failure because they have used reliable techniques (GC-MS, HPLC, etc.) to study the levels of CMPF in uremic/ dialyzed patients. On the basis of these findings, we can conclude that CMPF is responsible for renal failure by getting accumulated in the renal cells and due to its prooxidant nature, it increases the ROS production in the endothelial cells, which ultimately leads to the renal cellular damage. Another role of CMPF in renal damage which may not be denied is its ability to bind very strongly to HSA making itself a strong potent inhibitor resulting in accumulation and ultimately to renal failure. It may be of great importance to remove accumulated CMPF from HSA to prevent renal failure due to its inhibiting properties. Other compounds cannot easily replace CMPF which has a higher binding affinity. Addressing this major problem, we have shown that, DTQ - a compound which is already well- known for its beneficial effects on human health, and an active component of black cumin seed (*Nigella sativa*) has higher binding affinity towards HSA than the CMPF and higher lipophilicity due to which it may be proven as a potential drug candidate in curing renal failure caused by CMPF.

## Black Cumin Seed (*Nigella sativa*)

CMPF being the major drug-binding inhibitor, its binding with HSA could be restricted using any other compound with higher binding affinity and high therapeutic efficacy. Black cumin seed (*Nigella sativa*) has been known for its therapeutic efficiencies since many centuries. It has been referred in Islamic tradition as having healing powers [51]. It is a spice grown in Mediterranean region and in Western Asian countries such as India, Afghanistan, and Pakistan. The historical references to the seeds are also found in some of the oldest religious medical texts such as it is referred by Hippocrates and Dioscorides as 'Melanthion', the Bible describes it as the curative black cumin (Isaiah 28:25, 27 NKJV), and according to hadith, Islamic messenger Prophet Muhammad (PBUH) said "In the black seed is healing for every disease except death" (Sahih Bukhari).

Black cumin seeds have also been used in traditional medicines for many past years for the treatment of a wide range of diseases including a headache, bronchial asthma, infections, dysentery, back pain, hypertension, and gastrointestinal problems [52]. The pharmacological investigations of black cumin seed extracts have revealed its potential broad spectrum activities as antihistaminic [53], antimicrobial [54], anti-diabetic [55], anti-hypertensive [56], immunopotentiation [57], and anti-inflammatory [58]. It has also been proved to be consisting of anti-tumor [12] and anticancer activities [59-61].

## DTQ- a solution to CMPF accumulation problem

The bioactive constituents of the volatile black seed oil were first identified by Al-Dakhakhany (1963) [62] showing thymoquinone (TQ) as the most active constituent (about 54%) among the others. Later on, other active compounds: DTQ, thymol (THY), and thymohydroquinone (THQ) were identified by Ghosheh et al., (1999) [11] using HPLC. TQ (2-isopropyl-5-methyl-

1,4-benzoquinone) has been shown to exhibit anti-inflammatory, anti-oxidant, anti-neoplastic activities both *in-vivo* and *in-vitro* [63]. DTQ, THY, and THQ are more likely the metabolites of TQ as under physiological conditions, TQ gets slowly reduced to DTQ and THY [64].

TQ has been shown to possess antioxidant properties through different mechanisms such as it inhibits the formation of 5-hydroxyeicosa-tetraenaoic as well as 5-lipoxygenase products [65] which are required for the viability of colon cancer cells. Most importantly, in the context of this study, TQ has been shown to act as a scavenger to various ROS [66-68]. The molecular surfaces of TQ and DTQ possess significant amount of positively charged electron-deficit regions so that they may be subjected to nucleophilic attacks by glutathione and nucleotide bases in DNA, thereby, causing cellular toxicity due to the depletion of glutathione and damage of DNA due to the oxidation of nucleotide bases.

Badary et al., (1997) [69] studied the effects of TQ on cisplatin-induced nephrotoxicity and concluded that TQ ameliorates the nephrotoxicity in rodents and induce the anti-tumor activity of cisplatin. Due to the antioxidative and anti-inflammatory properties of TQ, it acts against renal injuries [15]. Fouda et al., (2008) [70] studied the effects of TQ on renal oxidative damage and proliferative response induced by mercuric chloride in rats and based on their results it was concluded that it may be a clinically valuable agent in the prevention of acute renal failure caused by inorganic mercury intoxication. Similarly, Hamed et al., (2013) [17] studied the effects of black cumin seed oil on hepato-renal injury induced by bromobenzene exposure. They concluded that the treatment with black seed oil significantly enhanced the hepato-renal protection mechanism, reduced disease complications, and delayed its progression [17].

All these investigations done earlier supports that renal failure/damage can be prevented/treated with black cumin seed due to the presence of anti-oxidative component TQ

and/ DTQ. TQ and DTQ both are similar in a way that they both have oxygen atoms on 1,4-positions of the benzene ring (Fig. 1), which contribute to their anti-oxidative properties. In this study, we performed docking experiments on HSA with DTQ (having same properties as TQ) and HSA with CMPF and found that DTQ was able to bind more effectively than CMPF.

## Therapeutics available for uremic ailments

Numerous uremic toxins such as indoles and phenols were detected for the first time in uremic plasma between the 1940s and 1990s [71-73]. It has been proved that a majority of uremic toxins originate endogenously through mammalian metabolism and therefore, it is being increasingly recognized that intestinal microbial metabolism has a large effect on mammalian blood metabolites resulting in the formation of various uremic toxins such as indoxyl sulphate, p-cresol sulphate, etc. [74-76]. CMPF, which is also a uremic toxin, is said to a metabolite of F-acids [36-40] and there are still no evidence to locate the place of its metabolization in humans. There are two main existing therapeutic approaches to treat ailment caused by uremic toxins: 1) interventions that modulate bacterial growth such as probiotics, prebiotics, and dietary modifications. 2) adsorbents that bind uremic toxins to reduce their absorption by the host. Both of these approaches are based on the fact that these uremic toxins are metabolized by the gut microbiota, but in the case of CMPF, it is still rendered questionable. However, the adsorbent approach is sometimes not helpful in retarding the progression of chronic kidney disease, for example, Sevelamer hydrochloride (Renagel; Genzyme, Cambridge, MA, USA) has been shown to bind some of the uremic toxins such as indoxyl sulphate (10-15%) and p-cresol (40-50%, depending on pH) [77], but it did not result in decreased serum concentrations of these uremic toxins in a mouse model of chronic kidney disease. However, the treatment of sevelamer in humans for studying the alterations in serum concentrations of microbial metabolites remains to be investigated.

# METHODS

## Retrieval of the structures

The 3D structures of DTQ and CMPF were downloaded from PubChem (PubChem CID: 398941 and PubChem CID: 123979, respectively). The protein crystal structure of HSA was downloaded from Protein Data Bank (PDB ID: 2BXA) [78].

## Molecular docking simulations

As revealed by Ghuman et al., (2005) [78], CMPF binds to the Y-150, R-257, H-242, R-222, and K-199 in the site I of HSA. Molecular docking simulations were done by AutoDock Vina [19]. CMPF and DTQ were docked with HSA by having default x, y, and z coordinates, i.e., 40 for each, and the center coordinates were set accordingly. The docking results were analyzed using PyMol [79].

## Binding Analysis

Some of the factors which were influenced by binding of CMPF and DTQ with HSA were analyzed by BINANA software [80]. BINANA (BINding ANAlyzer) is a python-implemented algorithm to analyze ligand binding. This software analyzes the key binding characteristics such as pi-pi interactions, hydrogen bonds, salt bridges, etc. In this study, we used BINANA to identify the active-site flexibility of CMPF and DTQ with HSA. The results are shown in Table 1.

## ADME properties analysis

The ADME properties of DTQ were predicted by SwissADME software [20]. The ADME properties predicted by the SwissADME include physicochemical properties, lipophilicity, water solubility, pharmacokinetics, drug-likeness, and medicinal chemistry. These properties are shown in Table 2.

# RESULTS

## Docking simulation results

Docking results showed that DTQ bound more effectively than CMPF showing the higher binding affinity (-8.1 kcal/mol) than the later (-7.0 kcal/mol). DTQ was able to occupy the binding pocket more efficiently than the CMPF (Fig. 2A and 2B). Docking simulations of CMPF with HSA showed the interaction with four residues of HSA: Y150, K199, R222, and R257 having the bond lengths equal to 2.9 A, 3.0 A, 3.0 A, and 3.2 A respectively (Fig. 3A). Similarly, DTQ bound with Y150, K199, R222, and R257 residues of HSA having the bond lengths equal to 3.0 A, 2.7 A, 3.0 A, and 3.2 A respectively (Fig. 3B). The bond lengths formed by DTQ are comparatively smaller than that of CMPF, this is the another parameter to support that DTQ is capable of binding more efficiently than CMPF and since the former bound at the same position as the later one, we can conclude that DTQ may be a potent inhibitor of CMPF.

**Fig. 2** Binding pocket of HSA **A)** CMPF **B)** DTQ. DTQ bind in the same pocket as CMPF in HSA and covers the binding site more efficiently than the CMPF.

**Fig. 3** Representing the interaction and bond lengths of **A)** CMPF and **B)** DTQ with HSA. DTQ showed shorter bond lengths with the interacting residues than that of CMPF.

## Binding Analysis results

According to Koshland, the active-site residues usually adjust to allow the binding of a specific substrate [81]. The induced-fit model assumes that the active site is flexible and changes shape to allow the complete binding of a substrate [82], which is a case here for CMPF and DTQ binding with HSA. BINANA categorizes each atom into six possible characterizations: alpha-sidechain, alpha-backbone, beta- sidechain, beta- backbone, other-sidechain, and other-backbone. The number of close- contact receptor atoms falling under each of these six categories is tallied as a metric of binding-site flexibility.  According to the predicted results of BINANA, DTQ showed more flexibility than CMPF towards the backbone and the side-chain (Table 1). These results suggest that the binding residues of HSA have to undergo more conformational changes for DTQ which may be due to the more complex structure than that of CMPF, and it may also contribute to the higher binding affinity of the former than the later.

**Table 1** Binding pocket flexibility properties of CMPF and DTQ.

| Ligand | Side-chain/Backbone | Secondary structure | Count |
|--------|---------------------|---------------------|-------|
| CMPF | Side-chain | Alpha | 153 |
| | Backbone | Alpha | 111 |
| DTQ | Side-chain | Alpha | 208 |
| | Backbone | Alpha | 153 |
| | Side-chain | Other | 4 |

## ADME properties

The ADME properties of DTQ were predicted in order to determine its lipophilicity.  Lipophilicity is one of the important parameters which is considered for drug candidates metabolism, greater the lipophilicity, more is the absorption. The ADME results are shown in Table 2.

**Table 2** ADME properties of CMPF and DTQ predicted by SwissADME.

| ADME properties | DTQ |
|-----------------|-----|
| **Molecular weight** | 328.40 g/mol |
| **Lipophilicity (WLogP)** | 2.71 |
| **Consensus Log $P_{o/w}$** | 2.62 |
| **Water Solubility (ESOL)** | -3.05 |
| **GI absorption** | High |
| **BBB permeant** | Yes |
| **Log $K_p$ (skin permeation)** | -6.83 cm/s |
| **Lipinski rule** | Yes |
| **Leadlikeness** | Yes |
| **Synthetic accessibility** | 4.65 |

Lipophilicity is the ability of the compound to dissolve in lipophilic solutions to permeate through the various biological membranes. It is measured the distribution of the compound between the non-aqueous (octanol) and the aqueous (water) phase, and the result is expressed as the logarithm of the concentration ratios termed as logP. A desired logP value is less than 5.0 [83]. In the case of DTQ, the estimated logP value is 2.62, which shows its higher lipophilicity. The BBB permeability is not considered good for the compounds as it may cause other harmful effects to the brain, but in the case of black cumin seed, it has been found that it causes beneficial neuropsychiatric effects and helps in to improve memory, and mood [84-86].

## BOILED-Egg results predicted by SwissADME

Brain Or IntestinaL EstimateD permeation (BOILED-Egg) method predicts an accurate model to analyze the gastrointestinal (GI) absorption and blood-brain barrier (BBB) of the candidate drugs [87]. It works by calculating the lipophilicity and polarity of the molecules. The BOILED-Egg representation of DTQ is shown in Fig. 4. The white region represents the physicochemical space of the molecule possessing the highest probability of absorption by the GI tract, whereas the yellow region represents the physicochemical space of the molecule having the highest probability to permeate the brain. According to the BOILED-Egg representation generated by the SwissADME, DTQ has high GI absorption and the BBB permeability (Fig. 4).

## CONCLUSION AND FUTURE DEVELOPMENTS

In this study we attempted to provide an insight into the role of CMPF in renal failure as implicated by many studies since quiet a time. CMPF has been found to be involved in causing type 2 diabetes by disrupting the beta-cell function [88-91], but it stands contradictory as some of the research does not correlate CMPF with diabetes [92,93] and also due to the lack of evidences CMPF has not been positively correlated with diabetes [94]. Therefore, we have addressed only the problem of CMPF binding with HSA leading to its accumulation and causing renal failure. According to the studies mentioned above, the capability of CMPF to bind strongly with HSA has posed a difficult task to remove the accumulated CMPF from the renal tubular cells. In this paper, we have proposed a suitable solution for the situation through *in silico* studies. We conclude that DTQ is capable to bind more strongly to HSA than the CMPF and shows inhibitory effects, which requires further *in vitro* studies to elucidate its actual mechanism of inhibition. The past publications provide sufficient evidences regarding the health benefits of active components of black cumin seed (*Nigella sativa*), therefore, we can conclude that DTQ possess many health benefits for humans due to its capability to bind with HSA more efficiently than CMPF, making it the best possible solution for inhibition of CMPF binding to HSA. Also, as CMPF is a strong lipophilic uremic solute (Depner, 1981), therefore, we comparatively analyzed the ADME properties of DTQ and CMPF *in silico*. According to the predicted ADME properties, DTQ could be considered as the potential drug candidate against the action of CMPF and can be further utilised to make therapeutic drugs to treat renal failure caused by the accumulation of CMPF in renal cells.

## Conflict of Interest

Authors declare that there is no conflict of interest whatsoever.

## Declaration

MF and TA conceived the idea and performed the experiments. MF, TA and YW wrote the manuscript.

# REFERENCES

1.  Mabuchi, H., & Nakahashi, H. (1988a). Inhibition of hepatic glutathione S-transferases by a major endogenous ligand substance present in uremic serum. *Nephron*, *49*(4), 281-283.

2.  Mabuchi, H., & Nakahashi, H. (1988b). A major endogenous ligand substance involved in renal failure. *Nephron*, *49*(4), 277-280.

3.  Mabuchi, H., & Nakahashi, H. (1990). Endogenous ligands that bind to serum albumin and renal failure. *Nephron*, *55*(1), 81-82.

4.  Sato, M., Koyama, M., Miyazaki, T., & Niwa, T. (1996). Reduced renal clearance of furancarboxylic acid, a major albumin-bound organic acid, in undialyzed uremic patients. *Nephron*, *74*(2), 419-421.

5.  Niwa, T. (1996, May). Organic acids and the uremic syndrome: protein metabolite hypothesis in the progression of chronic renal failure. In *Seminars in nephrology* (Vol. 16, No. 3, pp. 167-182).

6. Mabuchi, H., & Nakahashi, H. (1988c). A major inhibitor of phenytoin binding to serum protein in uremia. *Nephron*, *48*(4), 310-314.

7. Itoh, Y., Ezawa, A., Kikuchi, K., Tsuruta, Y., & Niwa, T. (2012). Protein-bound uremic toxins in hemodialysis patients measured by liquid chromatography/tandem mass spectrometry and their effects on endothelial ROS production. *Analytical and bioanalytical chemistry*, *403*(7), 1841-1850.

8. Mabuchi, H., & Nakahashi, H. (1987). Underestimation of serum albumin by the bromcresol purple method and a major endogenous ligand in uremia. *Clinica chimica acta*, *167*(1), 89-96.

9. Niwa, T. (1996, May). Organic acids and the uremic syndrome: protein metabolite hypothesis in the progression of chronic renal failure. In *Seminars in nephrology* (Vol. 16, No. 3, pp. 167-182).

10. Sarnatskaya, V. V., Lindup, W. E., Niwa, T., Ivanov, A. I., Yushko, L. A., Tjia, J., ... & Nikolaev, V. G. (2002). Effect of protein-bound uraemic toxins on the thermodynamic characteristics of human albumin. *Biochemical pharmacology*, *63*(7), 1287-1296.

11. Ghosheh, O. A., Houdi, A. A., & Crooks, P. A. (1999). High performance liquid chromatographic analysis of the pharmacologically active quinones and related compounds in the oil of the black seed (Nigella sativa L.). *Journal of pharmaceutical and biomedical analysis*, *19*(5), 757-762.

12. Ait Mbarek, L., Ait Mouse, H., Elabbadi, N., Bensalah, M., Gamouh, A., Aboufatima, R., ... & Zyad, A. (2007). Anti-tumor properties of blackseed (Nigella sativa L.) extracts. *Brazilian Journal of Medical and Biological Research*, *40*(6), 839-847.

13. Padhye, S., Banerjee, S., Ahmad, A., Mohammad, R., & Sarkar, F. H. (2008). From here to eternity-the secret of Pharaohs: Therapeutic potential of black cumin seeds and beyond. *Cancer therapy*, *6*(b), 495.

14. Sultan, M. T., Butt, M. S., Anjum, F. M., Jamil, A., Akhtar, S., & Nasir, M. (2009). Nutritional profile of indigenous cultivar of black cumin seeds and antioxidant potential of its fixed and essential oil. *Pak J Bot*, *41*(3), 1321-1330.

15. Ragheb, A., Attia, A., Eldin, W. S., Elbarbry, F., Gazarin, S., & Shoker, A. (2009). The protective effect of thymoquinone, an anti-oxidant and anti-inflammatory agent, against renal injury: a review. *Saudi Journal of Kidney Diseases and Transplantation*, *20*(5), 741.

16. Huq, F., & Mazumder, E. H. (2010). Molecular modelling analysis of the metabolism of thymoquinone.

17. Hamed, M. A., & Ali, S. A. (2013). Effects of black seed oil on resolution of hepato-renal toxicity induced by bromobenzene in rats. *Eur Rev Med Pharmacol Sci*, *17*(5), 569-81.

18. Mathur, M. L., Gaur, J., Sharma, R., & Haldiya, K. R. (2011). Antidiabetic properties of a spice plant Nigella sativa. *Journal of Endocrinology and Metabolism*, *1*(1), 1-8.

19. Trott, O., & Olson, A. J. (2010). AutoDock Vina: improving the speed and accuracy of docking with a new scoring function, efficient optimization, and multithreading. *Journal of computational chemistry*, *31*(2), 455-461.

20. Daina, A., Michielin, O., & Zoete, V. (2017). SwissADME: a free web tool to evaluate pharmacokinetics, drug-likeness and medicinal chemistry friendliness of small molecules. *Scientific Reports*, *7*, 42717.

21. Peters, T. (1995). All About Albumin: Biochemistry, Genetics and Medical Applications, Academic Press, San Diego.

22. Curry, S., Mandelkow, H., Brick, P. & Franks, N. (1998). Crystal structure of human serum albumin complexed with fatty acid reveals an asymmetric distribution of binding sites. Nature Struct. Biol. 5, 827–835.

23. Bhattacharya, A. A., Gru¨ne, T. & Curry, S. (2000). Crystallographic analysis reveals common modes of binding of medium and long-chain fatty acids to human serum albumin. J. Mol. Biol. 303, 721–732

24. Petitpas, I., Gru¨ne, T., Bhattacharya, A. A. & Curry, S. (2001). Crystal structures of human serum albumin complexed with monounsaturated and polyunsaturated fatty acids. J. Mol. Biol. 314, 955–960.

25. Wardell, M., Wang, Z., Ho, J. X., Robert, J., Ruker, F., Ruble, J. & Carter, D. C. (2002). The atomic structure of human methemalbumin at 1.9 A˚. Biochem. Biophys. Res. Commun. 291, 813–819.

26. Petitpas, I., Petersen, C. E., Ha, C. E., Bhattacharya, A. A., Zunszain, P. A., Ghuman, J. et al. (2003). Structural basis of albumin–thyroxine interactions and familial dysalbuminemic hyperthyroxinemia. Proc. Natl Acad. Sci. USA, 100, 6440–6445.

27. Zunszain, P. A., Ghuman, J., Komatsu, T., Tsuchida, E. & Curry, S. (2003). Crystal structural analysis of human serum albumin complexed with hemin and fatty acid. BMC Struct. Biol. 3, 6

28. He, X. M., & Carter, D. C. (1992). Atomic structure and chemistry of human serum albumin.

29. Sudlow, G.D.J.B., Birkett, D.J. and Wade, D.N., (1975). The characterization of two specific drug binding sites on human serum albumin. Molecular Pharmacology, 11(6), pp.824-832.

30. Müller, W. E., & Wollert, U. (1979). Human serum albumin as a 'silent receptor' for drugs and endogenous substances. *Pharmacology*, *19*(2), 59-67.

31. Lejon, S., Frick, I. M., Björck, L., Wikström, M., & Svensson, S. (2004). Crystal structure and biological implications of a bacterial albumin binding module in complex with human serum albumin. *Journal of biological chemistry*, *279*(41), 42924-42928.

32. Fasano, M., Curry, S., Terreno, E., Galliano, M., Fanali, G., Narciso, P., ... & Ascenzi, P. (2005). The extraordinary ligand binding properties of human serum albumin. *IUBMB life*, *57*(12), 787-796.

33. Ascoli, G. A., Domenici, E., & Bertucci, C. (2006). Drug binding to human serum albumin: Abridged review of results obtained with high-performance liquid chromatography and circular dichroism. *Chirality*, *18*(9), 667-679.

34. Yang, F., Zhang, Y., & Liang, H. (2014). Interactive association of drugs binding to human serum albumin. *International journal of molecular sciences*, *15*(3), 3580-3595.

35. Hein, K. L., Kragh-Hansen, U., Morth, J. P., Jeppesen, M. D., Otzen, D., Møller, J. V., & Nissen, P. (2010). Crystallographic analysis reveals a unique lidocaine binding site on human serum albumin. *Journal of structural biology*, *171*(3), 353-360.

36. Schödel, R., Dietel, P., and Spiteller, G. (1986) F-Säuren als Vorstufen der Urofuransäuren, Liebigs Ann. Chem., 127–131.

37. Spiteller M, Spiteller G. (1979). Separation and characterization of acidic urine constituents (author's transl). Journal of Chromatography [1979, 164(3):253-317]. PMID:544607

38. Spiteller, M., Spiteller, G., & Hoyer, G. A. (1980). Urofuransäuren–eine bisher unbekannte Klasse von Stoffwechselprodukten. Chemische Berichte, 113(2), 699-709.

39. Pfordt, J., Thoma, H. and Spiteller, G., (1981). Identifizierung, Strukturableitung und Synthese bisher unbekannter Urofuransäuren im menschlichen Blut. *Liebigs Annalen der Chemie*, *1981*(12), pp.2298-2308.

40. Wahl, H. G., Tetschner, B., & Liebich, H. M. (1992). The effect of dietary fish oil supplementation on the concentration of 3-carboxy-4-methyl-5-propyl-2-furanpropionic acid in human blood and urine. Journal of High Resolution Chromatography, *15*(12), 815-818.

41. Costigan, M.G., and Lindup, W.E. (1996) Plasma Clearance in the Rat of Furan Dicarboxylic Acid Which Accumulates in Uremia, Kidney Int. 49, 634–638.

42. Deguchi, T., Kusuhara, H., Takadate, A., Endou, H., Otagiri, M. and Sugiyama, Y., (2004). Characterization of uremic toxin transport by organic anion transporters in the kidney. Kidney international, 65(1), pp.162-174.

43. Henderson, S.J., and Lindup, E. (1992) Renal Organic Acid Transport: Uptake by Rat Kidney Slices of a Furan Dicarboxylic Acid Which Inhibits Plasma Protein Binding of Acidic Ligands in Uremia, J. Pharmacol. Exp. Ther. 263, 54–60.

44. Sakai, Takadate A., Otagiri M. (1995). Characterization of binding site of uremic toxins on human serum albumin [J]. Biological and Pharmaceutical Bulletin. 18(12): 1755-1761.

45. Tsutsumi Y, Deguchi T, Takano M, et al (2002). Renal disposition of a furan dicarboxylic acid and other uremic toxins in the rat[J]. Journal of Pharmacology and Experimental Therapeutics. 303(2): 880-887.

46. Lim, C. F., Bernard, B. F., De Jong, M., Docter, R., Krenning, E. P., & Hennemann, G. (1993). A furan fatty acid and indoxyl sulfate are the putative inhibitors of thyroxine hepatocyte transport in uremia. *The Journal of Clinical Endocrinology & Metabolism*, *76*(2), 318-324.

47. Franke, R. M., & Sparreboom, A. (2008). Inhibition of imatinib transport by uremic toxins during renal failure. *Journal of Clinical Oncology*, *26*(25), 4226-4227.

48. Motojima, M., Hosokawa, A., Yamato, H., Muraki, T., & Yoshioka, T. (2003). Uremic toxins of organic anions up-regulate PAI-1 expression by induction of NF-κB and free radical in proximal tubular cells. *Kidney international*, *63*(5), 1671-1680.

49. Shimoishi, K., Anraku, M., Kitamura, K., Tasaki, Y., Taguchi, K., Hashimoto, M., ... & Otagiri, M. (2007). An oral adsorbent, AST-120 protects against the progression of oxidative stress by reducing the accumulation of indoxyl sulfate in the systemic circulation in renal failure. *Pharmaceutical research*, *24*(7), 1283-1289.

50. Miyamoto Y, Iwao Y, Mera K, et al (2012). A uremic toxin, 3-carboxy-4-methyl-5-propyl-2-furanpropionate induces cell damage to proximal tubular cells via the generation of a radical intermediate [J]. Biochemical pharmacology. 84(9): 1207-1214.

51. Goreja, W. G. (2003). Black Seed: Natural Medical Remedy. *Amazing Herbs*.

52. Al-Rowais, N. A. (2002). Herbal medicine in the treatment of diabetes mellitus. *Saudi medical journal*, *23*(11), 1327-1331.

53. Mahfouz, M., ABDELMAG. R, & ELDAKHAK. M. (1965). EFFECT OF NIGELLONE-THERAPY ON HISTAMINOPEXIC POWER OF BLOOD SERA OF ASTHMATIC PATIENTS. *Arzneimittel-forschung*, *15*(10), 1230.

54. El-Fatatry, H. M. (1975). Isolation and structure assignment of an antimicrobial principle from the volatile oil of Nigella sativa L. seeds. *Die Pharmazie*, *30*(2), 109-111.

55. Al-Hader, A., Aqel, M., & Hasan, Z. (1993). Hypoglycemic effects of the volatile oil of Nigella sativa seeds. *International journal of pharmacognosy*, *31*(2), 96-100.

56. El Tahir, K. E., Ashour, M. M., & Al-Harbi, M. M. (1993). The cardiovascular actions of the volatile oil of the black seed (Nigella sativa) in rats: elucidation of the mechanism of action. *General Pharmacology: The Vascular System*, *24*(5), 1123-1131.

57. Medinica, R., Mukerjee, S., Huschart, T., & Corbitt, W. (1994, April). Immunomodulatory and anticancer activity of Nigella sativa plant extract in humans. In *Proceedings of the American Association for Cancer Research Annual Meeting* (p. A2865).

58. Houghton, P. J., Zarka, R., de las Heras, B., & Hoult, J. R. S. (1995). Fixed oil of Nigella sativa and derived thymoquinone inhibit eicosanoid generation in leukocytes and membrane lipid peroxidation. *Planta medica*, *61*(01), 33-36.

59. Khan, A., Chen, H. C., Tania, M., & Zhang, D. Z. (2011). Anticancer activities of Nigella sativa (black cumin). *African Journal of Traditional, Complementary and Alternative Medicines*, *8*(5S).

60. Harzallah, H. J., Kouidhi, B., Flamini, G., Bakhrouf, A., & Mahjoub, T. (2011). Chemical composition, antimicrobial potential against cariogenic bacteria and cytotoxic activity of Tunisian Nigella sativa essential oil and thymoquinone. *Food Chemistry*, *129*(4), 1469-1474.

61. Randhawa, M. A., & Alghamdi, M. S. (2011). Anticancer activity of Nigella sativa (black seed)—a review. *The American journal of Chinese medicine*, *39*(06), 1075-1091.

62. Al-Dakhakhany, M. (1963). Studies on the chemical constituition of Egyptian Nigella sativa L seeds. *Planta Med*, *1*, 465-470.

63. Gali-Muhtasib, H., Roessner, A., & Schneider-Stock, R. (2006). Thymoquinone: a promising anti-cancer drug from natural sources. *The international journal of biochemistry & cell biology*, *38*(8), 1249-1253.

64. Khalife, K. H., & Lupidi, G. (2007). Nonenzymatic reduction of thymoquinone in physiological conditions. *Free radical research*, *41*(2), 153-161.

65. El-Dakhakhny, M., Madi, N. J., Lembert, N., & Ammon, H. P. T. (2002). Nigella sativa oil, nigellone and derived thymoquinone inhibit synthesis of 5-lipoxygenase products in polymorphonuclear leukocytes from rats. *Journal of ethnopharmacology*, *81*(2), 161-164.

66. Kruk, I., Michalska, T., Lichszteld, K., Kładna, A., & Aboul-Enein, H. Y. (2000). The effect of thymol and its derivatives on reactions generating reactive oxygen species. *Chemosphere*, *41*(7), 1059-1064.

67. Mansour, M. A., Nagi, M. N., El-Khatib, A. S., & Al-Bekairi, A. M. (2002). Effects of thymoquinone on antioxidant enzyme activities, lipid peroxidation and DT-diaphorase in different tissues of mice: a possible mechanism of action. *Cell biochemistry and function*, *20*(2), 143-151.

68. Badary, O. A., Taha, R. A., Gamal El-Din, A. M., & Abdel-Wahab, M. H. (2003). Thymoquinone is a potent superoxide anion scavenger. *Drug and chemical toxicology*, *26*(2), 87-98.

69. Badary, O. A., Nagi, M. N., Al-Shabanah, O. A., Al-Sawaf, H. A., Al-Sohaibani, M. O., & Al-Bekairi, A. M. (1997). Thymoquinone ameliorates the nephrotoxicity induced by cisplatin in rodents and potentiates its antitumor activity. *Canadian Journal of Physiology and Pharmacology*, *75*(12), 1356-1361.

70. Fouda, A. M. M., Daba, M. H. Y., Dahab, G. M., & Sharaf el-Din, O. A. (2008). Thymoquinone ameliorates renal oxidative damage and proliferative response induced by mercuric chloride in rats. *Basic & clinical pharmacology & toxicology*, *103*(2), 109-118.

71. LUDWIG, G. D., SENESKY, D., Bluemlef, L. W., & ELKINTON, J. R. (1968). Indoles in uremia: identification by countercurrent distribution and paper chromatography. *The American journal of clinical nutrition*, *21*(5), 436-450.

72. Toshimitsu, N., Kenji, M., Toyokazu, O., Akira, S., & Kaizo, K. (1981). A gas chromatographic-mass spectrometric analysis for phenols in uremic serum. *Clinica Chimica Acta*, *110*(1), 51-57.

73. Van Haard, P. M., & Pavel, S. (1988). Chromatography of urinary indole derivatives. *Journal of Chromatography B: Biomedical Sciences and Applications*, *429*, 59-94.

74. Wikoff, W. R., Anfora, A. T., Liu, J., Schultz, P. G., Lesley, S. A., Peters, E. C., & Siuzdak, G. (2009). Metabolomics analysis reveals large effects of gut microflora on mammalian blood metabolites. *Proceedings of the national academy of sciences*, *106*(10), 3698-3703.

75. Niwa, T. (2009). Recent progress in the analysis of uremic toxins by mass spectrometry. *Journal of Chromatography B*, *877*(25), 2600-2606.

76. Rhee, E. P., Souza, A., Farrell, L., Pollak, M. R., Lewis, G. D., Steele, D. J., ... & Gerszten, R. E. (2010). Metabolite profiling identifies markers of uremia. *Journal of the American Society of Nephrology*, *21*(6), 1041-2051.

77. De Smet, R., Thermote, F., Lameire, N., & Vanholder, R. (2004). Sevelamer Hydrochloride (renagel®) Adsorbs The Uremic Compound Indoxyl Sulfate. *The International Journal of Artificial Organs*, *27*(7), 569.

78. Ghuman, J., Zunszain, P. A., Petitpas, I., Bhattacharya, A. A., Otagiri, M., & Curry, S. (2005). Structural basis of the drug-binding specificity of human serum albumin. Journal of molecular biology, 353(1), 38-52

79. DeLano, W. L. (2002). The PyMOL molecular graphics system.

80. Durrant, J. D., & McCammon, J. A. (2011). BINANA: a novel algorithm for ligand-binding characterization. *Journal of Molecular Graphics and Modelling*, *29*(6), 888-893.

81. Koshland, D. E. (1959). Enzyme flexibility and enzyme action. *Journal of cellular and comparative physiology*, *54*(S1), 245-258.

82. Sullivan, S. M., & Holyoak, T. (2008). Enzymes with lid-gated active sites must operate by an induced fit mechanism instead of conformational selection. *Proceedings of the National Academy of Sciences*, *105*(37), 13829-13834.

83. Novamass- http://www.sbw.fi/lead-optimization/experimental-logp-logd-logs-pka-analysis/

84. Akhondian, J., Parsa, A., & Rakhshande, H. (2007). The effect of Nigella sativa L.(black cumin seed) on intractable pediatric seizures. *Medical Science Monitor*, *13*(12), CR555-CR559.

85. Alenazi, S. A. (2016). Neuropsychiatric Effects of Nigella sativa (Black Seed) A Review. *Alternative & Integrative Medicine*.

86. Beheshti, F., Khazaei, M., & Hosseini, M. (2016). Neuropharmacological effects of Nigella sativa. *Avicenna journal of phytomedicine*, *6*(1), 104.

87. Daina, A., & Zoete, V. (2016). A BOILED-Egg To Predict Gastrointestinal Absorption and Brain Penetration of Small Molecules. *ChemMedChem*, *11*(11), 1117-1121.

88. Prentice K J, Luu L, Allister E M, et al (2014). The furan fatty acid metabolite CMPF is elevated in diabetes and induces 尾 cell dysfunction[J]. Cell metabolism. 19(4): 653-666.

89. Liu Y, Prentice K J, Eversley J A, et al (2016). Rapid elevation in CMPF may act as a tipping point in diabetes development [J]. Cell reports. 14(12): 2889-2900.

90. Retnakaran R, Ye C, Kramer C K, et al. (2016). Evaluation of circulating determinants of beta-cell function in women with and without gestational diabetes [J]. The Journal of Clinical Endocrinology & Metabolism, 2016: jc. 2016-1402.

91. Lu Y, Wang Y, Ong C N, et al. (2016). Metabolic signatures and risk of type 2 diabetes in a Chinese population: an untargeted metabolomics study using both LC-MS and GC-MS[J]. Diabetologia. 59(11): 2349-2359.

92. Lankinen M A, Hanhineva K, Kolehmainen M, et al. (2015). CMPF does not associate with impaired glucose metabolism in individuals with features of metabolic syndrome[J]. PloS one 10(4): e0124379.

93. Zheng J S, Lin M, Imamura F, et al (2016). Serum metabolomics profiles in response to n-3 fatty acids in Chinese patients with type 2 diabetes: a double-blind randomised controlled trial[J]. Scientific Reports. 6.

94. Wallin, A., Di Giuseppe, D., Orsini, N., Patel, P.S., Forouhi, N.G. and Wolk, A., (2012). Fish consumption, dietary long-chain n-3 fatty acids, and risk of type 2 diabetes. Diabetes care, 35(4), pp.918-929.

## Figure legends

**Fig.1** 2D structure of **a)** TQ  **b)** DTQ

**Fig. 2** Binding pocket of HSA **A)** CMPF **B)** DTQ. DTQ bind in the same pocket as CMPF in HSA and covers the binding site more efficiently than the CMPF.

**Fig. 3** Representing the interaction and bond lengths of **A)** CMPF and **B)** DTQ with HSA. DTQ showed shorter bond lengths with the interacting residues than that of CMPF.

**Fig. 4** BOILED-Egg representation of DTQ. Yellow region represents the BBB and the white region represents the GI absorption, and red circle represents the molecule (i.e., DTQ).

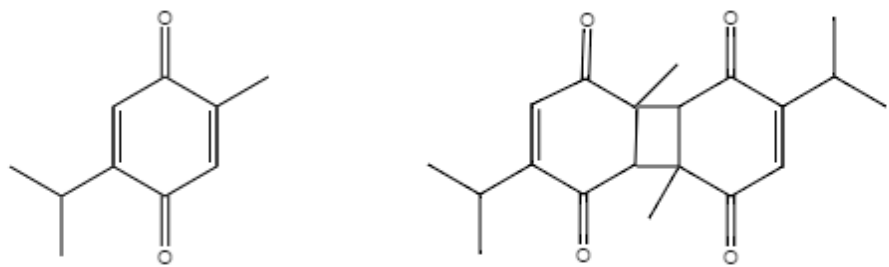

**Fig.1** 2D structure of **a)** TQ  **b)** DTQ

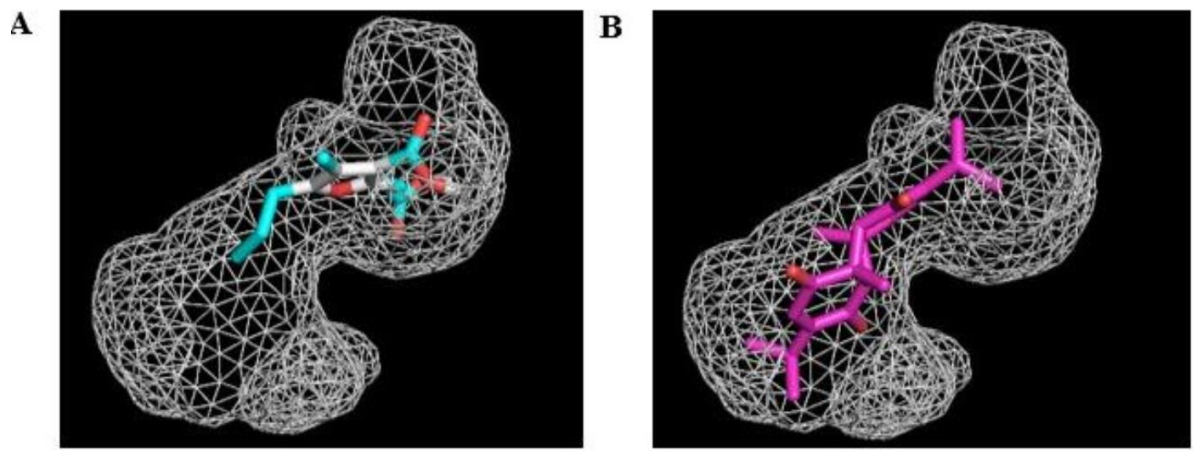

**Fig. 2** Binding pocket of HSA **A)** CMPF **B)** DTQ. DTQ bind in the same pocket as CMPF in HSA and covers the binding site more efficiently than the CMPF.

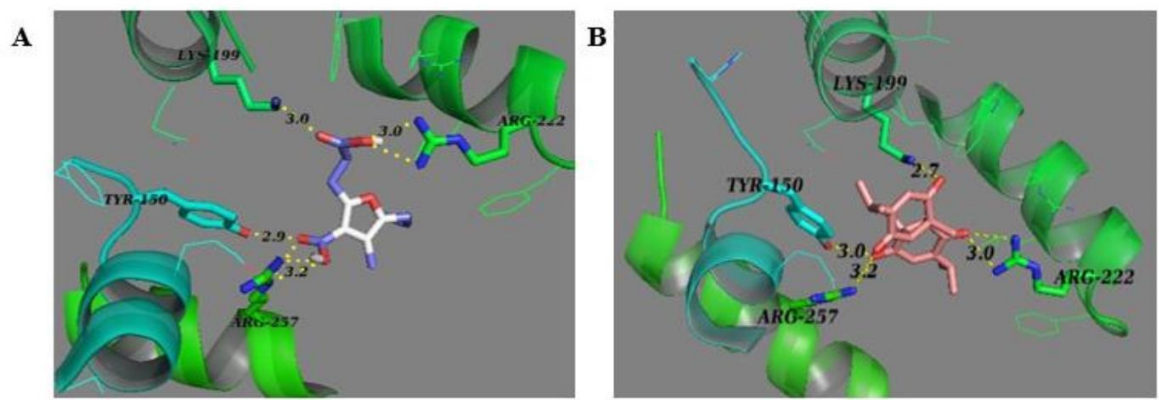

**Fig. 3** Representing the interaction and bond lengths of **A)** CMPF and **B)** DTQ with HSA. DTQ showed shorter bond lengths with the interacting residues than that of CMPF.

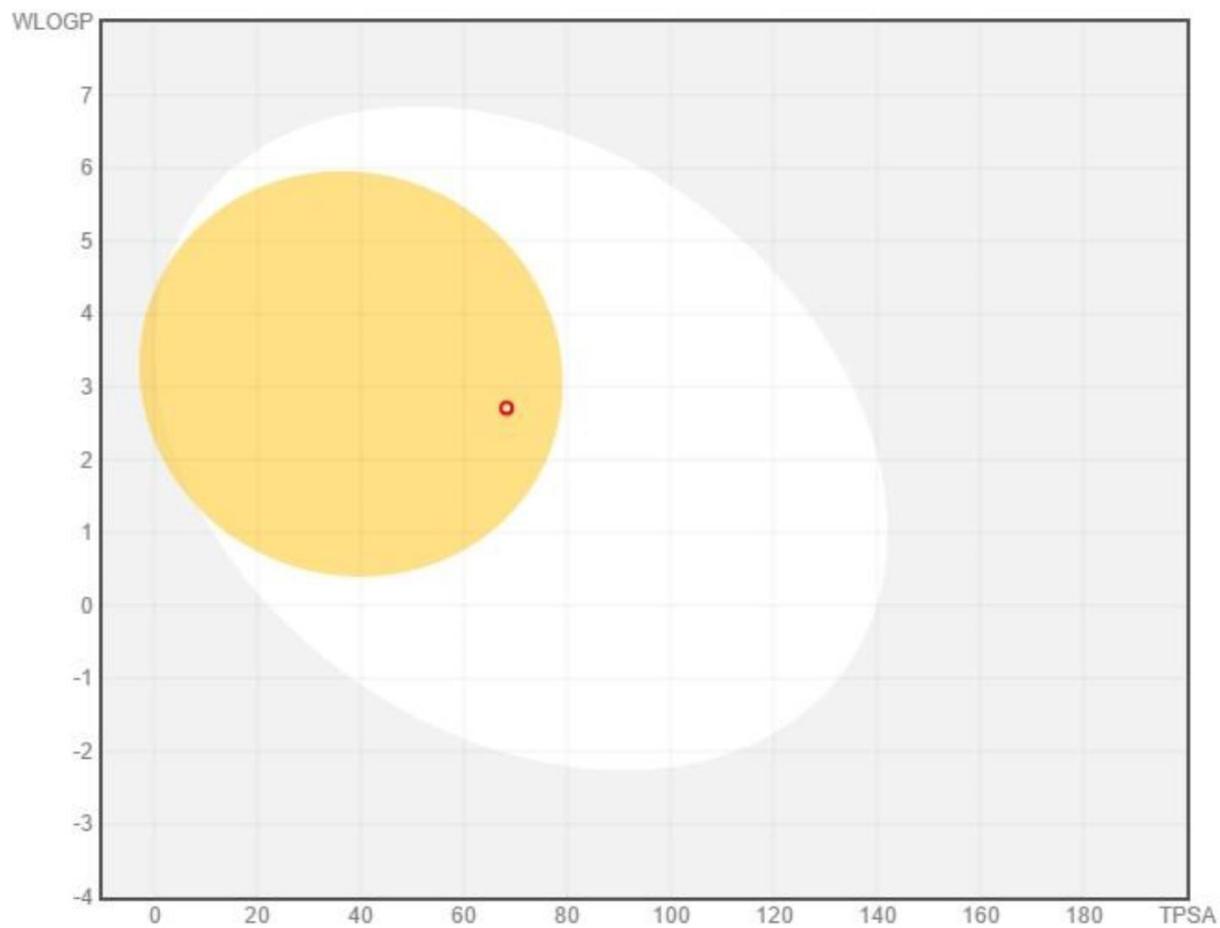

**Fig. 4** BOILED-Egg representation of DTQ. Yellow region represents the BBB and the white region represents the GI absorption, and red circle represents the molecule (i.e., DTQ).